\documentclass[3p]{elsarticle}

\global\preprintfalse
\usepackage{subfig}
\usepackage{graphicx}

\usepackage{textcomp}

\begin{document}

\title{Study of the performance of the NA62 Small-Angle Calorimeter at the DA$\Phi$NE Linac}
\author{A.~Antonelli$^a$}
\author{F.~Gonnella$^a$\footnote{Present address: University of Birmingham, 
School of Physics and Astronomy, Edgbaston, B152TT, Birmingham, West Midlands, UK}}
\author{V.~Kozhuharov$^{a,b}$ 
  }
\author{M.~Moulson$^a$}
\author{M.~Raggi$^a$\footnote{Present address: Sapienza Universita' di Roma, Piazzale Aldo Moro 5 Rome, Italy}}
\author{T.~Spadaro$^a$}
\address{$^a$Laboratori Nazionali di Frascati, 00044 Frascati RM, Italy}
\address{$^b$Faculty of Physics, University of Sofia ``St. Kl. Ohridski'', 5 J. Bourchier Blvd., 1164 Sofia, Bulgaria}

\begin{abstract}
The measurement of $BR(K^+\to\pi^+\nu\bar{\nu})$ with 10\% precision by the NA62 experiment 
requires extreme background suppression. 
The Small Angle Calorimeter aims to provide an efficient 
veto for photons flying at angles down to zero with respect to the 
kaon flight direction. The initial prototype was upgraded and tested at the Beam Test Facility 
of the DA$\Phi$NE Linac at Frascati. 
The energy resolution and the efficiency were measured and are presented.
\end{abstract}

\maketitle

\section{Introduction}

The NA62 experiment at CERN SPS aims to measure the 
branching ratio of the decay $K^+\to\pi^+\nu\bar{\nu}$ 
with $\sim10$\% precision \cite{bib:na62-TD}. 
The extremely small Standard Model (SM) prediction, 
$BR(K^+\to\pi^+\nu\bar{\nu}) = ( 8.4 \pm 1.0 )\times 10^{-11}$ \cite{bib:pinn-theory}, 
sets state-of-the-art requirements for the experimental setup
and analysis technique. 
To reject the background 
from kaon decay modes with photons in the final state 
a complex photon veto system consisting of 
Large Angle Vetoes (LAV), a Liquid Krypton Calorimeter (LKr), 
an Intermediate Ring Calorimeter (IRC), 
and a Small Angle Calorimeter (SAC) is used. 
The charged kaon beam is transported in vacuum till the end of the experimental complex where
a dipole magnet deflects it outside the acceptance of the Small Angle Calorimeter. 
The location of the SAC, downstream of the 
experimental setup 
on the axis of the charged beam before deflection
makes it exposed to photons with energy above 5 GeV \cite{bib:na62-TD}. 

\section{The Small-Angle calorimeter}

\begin{figure}[htb]
\begin{center}
\includegraphics[width=7cm]{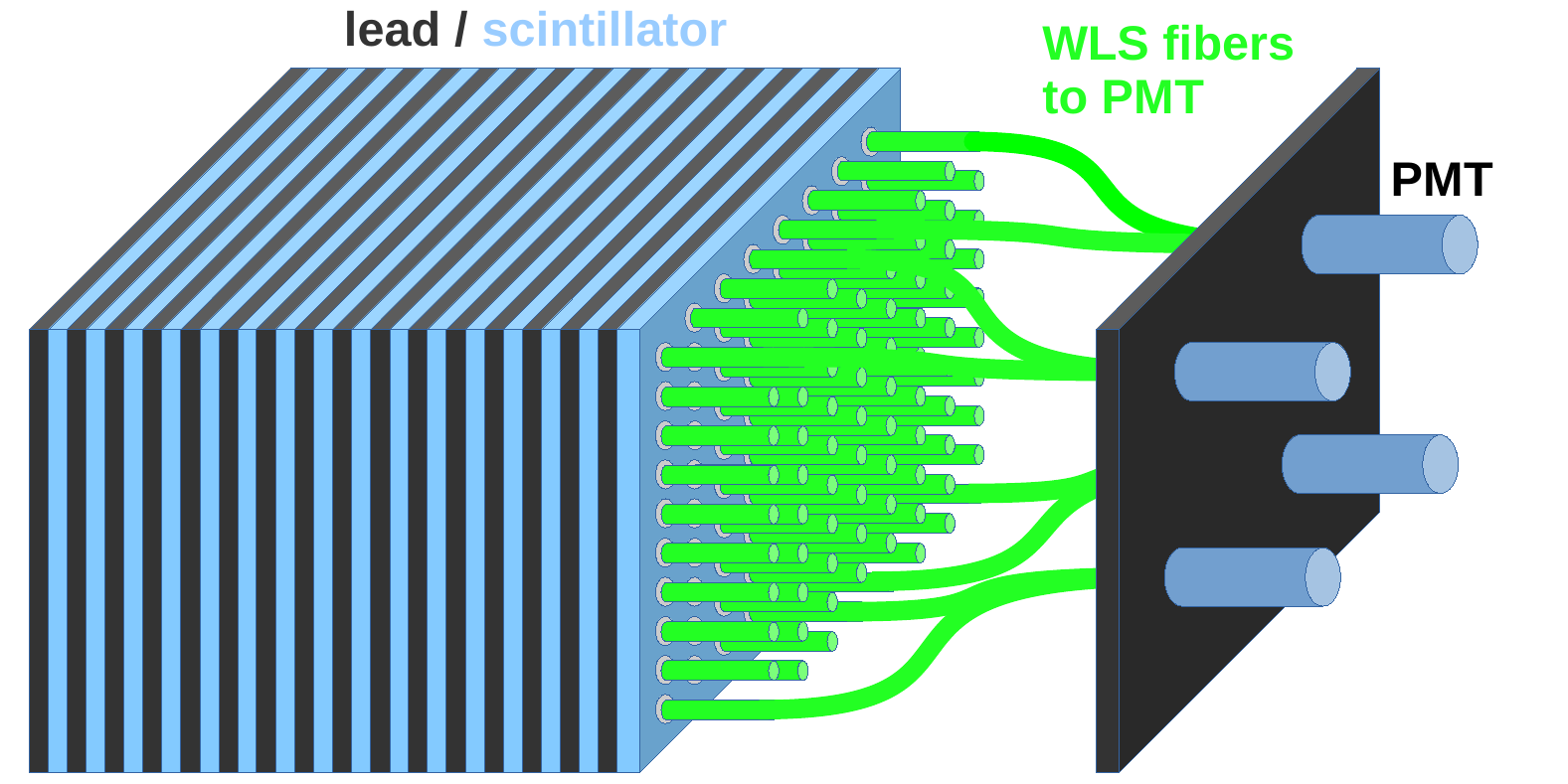}   
\end{center}
\caption{Schematic view of the SAC (not to scale) --
alternating plates of lead and scintillator, traversed by WLS fibers.}
\label{fig:sac-schem}
\end{figure}

A prototype of the Small Angle Calorimeter was 
constructed and tested at CERN during a 2006 test run \cite{bib:sac-construction}. 
It consisted of 70 lead plates and 70 scintillator plates, each with thickness
of 1.5 mm (Fig. \ref{fig:sac-schem}).
The light was collected by Kuraray 1 mm Y-11(250)MSJ Wave-Length Shifting (WLS) fibers
and was read out by four FEU-84 photomultipliers (PMT).  
Since the scintillator plates are monolithic the four readout channels
are optically connected and the SAC should be thought as a 
single channel detector, where the emitted light due to a single 
particle is shared among the four PMTs.
More details can be found elsewhere \cite{bib:sac-construction}. 


In 2012 a degradation of the PMT performance was noticed. 
The total collected charge for muons from cosmic rays
was an order of magnitude smaller than the expected.
The origin for this effect was traced back to possible improper storage conditions.
In addition, 
new estimates of
the particle rate expected at the SAC ($\sim 1$ MHz) 
required improvement of the
double pulse separation. This could be achieved by using fast PMTs
and/or changing the WLS fibers. 
It was decided to upgrade the prototype
by changing the photomultipliers to 25~mm Hamamatsu 
R6427 \cite{bib:hamamatsu-r6427}, which provide $\sim 1.7$ ns signal risetime. 

The upgraded prototype was shipped to the Frascati Beam Test  Facility (BTF) to test 
its performance.

\section{Test beam setup}

The Beam Test Facility of the DA$\Phi$NE linac at Frascati 
is a dedicated beam line able to deliver electrons or positrons in 
a separate hall. The beam can be used for detector testing purposes
and/or study of physics phenomena at the energy scale of O(100) MeV. 
The BTF can deliver up to 50 bunches of $e^+/e^-$ per second with energy 
in the range from 100 MeV up to 840 MeV for $e^-$ and 550 MeV for $e^+$ \cite{bib:btf}.
Once per second a bunch is sent to a beam monitor for feedback. 
The bunch length was adjusted to be 10 ns. 
An important feature of the facility is the possibility to vary the 
beam intensity from $\sim 10^9$ particles down to a single particle per bunch. 
The latter is extremely important for detector efficiency and energy resolution 
measurements. 
  The typical size of the beam spot was of the order of 1 cm$^2$ or less. 

\begin{figure}[htb]
\begin{center}
\includegraphics[width=8cm]{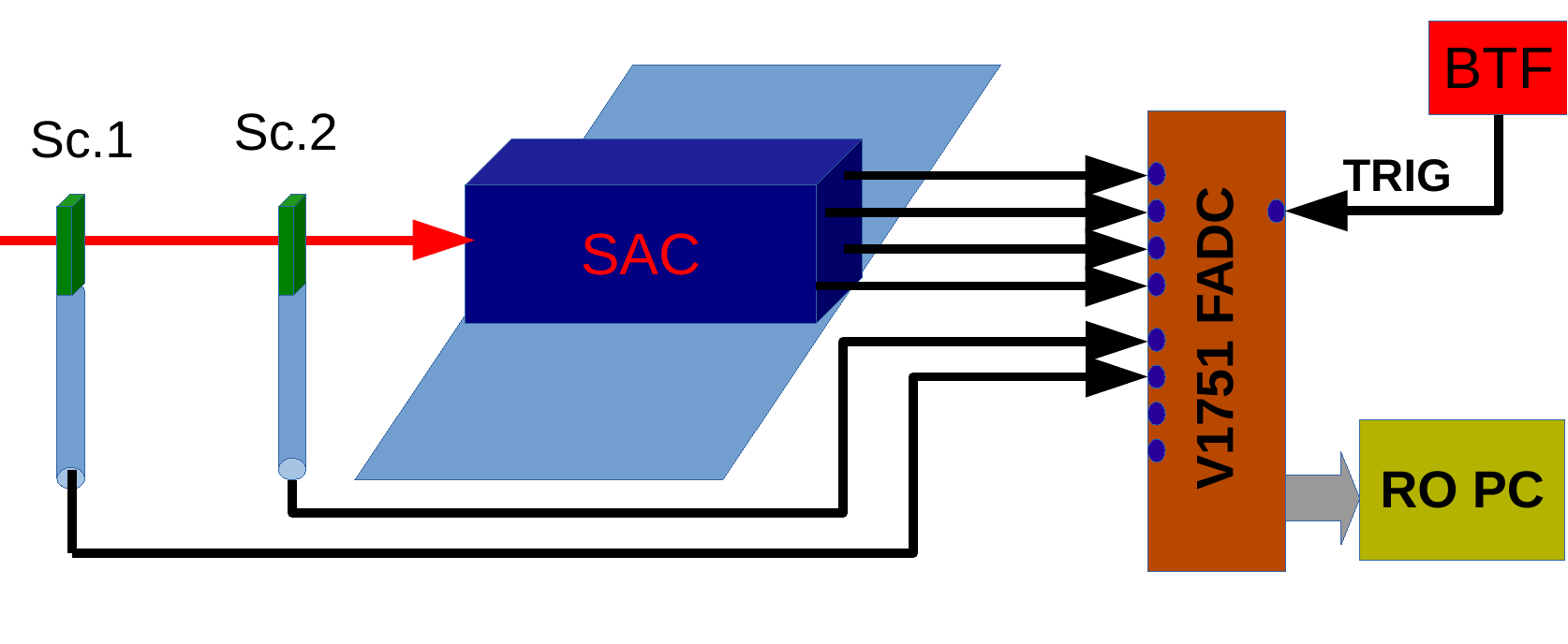}
\end{center}
\caption{Schematics of the test beam setup to study the SAC response to 606 MeV electrons at BTF, LNF-INFN.}
\label{fig:btf-setup}
\end{figure}

During the study of the response of the SAC 
the BTF was operated in electron mode and the average number of 
particles per single bunch was kept close to one. 
The SAC was placed on a movable table and was 
adjusted so that the BTF beam impinged on the detector close to its center. 
Two  paddles (Sc.1 and Sc.2) made of plastic scintillator with a thickness of 10 mm
 were placed in front of the SAC, 
as shown schematically in Fig. \ref{fig:btf-setup}. 
The same paddles had also been used during the measurement of the efficiency 
of the Large Angle Veto prototypes for the NA62 experiment \cite{bib:lav-prot-eff}. 
The SAC PMTs were operated at 900 V, corresponding to a gain $\sim 1.6\times10^5$. 
The four channels of the SAC and the signals from the scintillator paddles 
were fed into a CAEN V1751 FADC \cite{bib:caen-v1751}. The digitizer was operated in sampling mode, 
providing 10 bit amplitude measurement of the input signal amplitude at 1 GS per second. 
The trigger was based on an external signal from the BTF which  
allowed to record the waveforms of all the readout channels for every single bunch.
The lack of online event selection diminished the possible bias due to inefficiency
of a triggering system. 
The data was transferred to a readout PC through optical fibers.

The digitizer readout of the signals from Sc.1 and Sc.2 
provided the possibility to select events with a single electron, 
based on the signal amplitude and the collected charge by the paddle PMTs. 

The results presented  use the data from 684759 bunches collected during the test
run in the summer of 2013. The energy of the beam was set to 606 MeV
 and the detector performance was stable during the data taking.

\section{Energy resolution}

The offline data contained the 
recorded waveforms of readout signals
in a window of 1024 ns.
The data from one BTF bunch was written as a single event.  
The typical signal from the PMTs was contained within $\sim$ 40 ns, 
with a maximum positioned at about 330 ns from the start of the 
readout window. 
For each recorded event 
the individual PMT charges were obtained by integrating the 
recorded waveforms after pedestal subtraction.
The pedestal was determined on event by event basis using the amplitude measuments preceeding the 
trigger, to compensate for possible pedestal fluctuations. 
The total charge in the SAC was calculated as the sum 
of the charges in the four PMTs.

\begin{figure}[htb]
\begin{center}
\includegraphics[width=9cm]{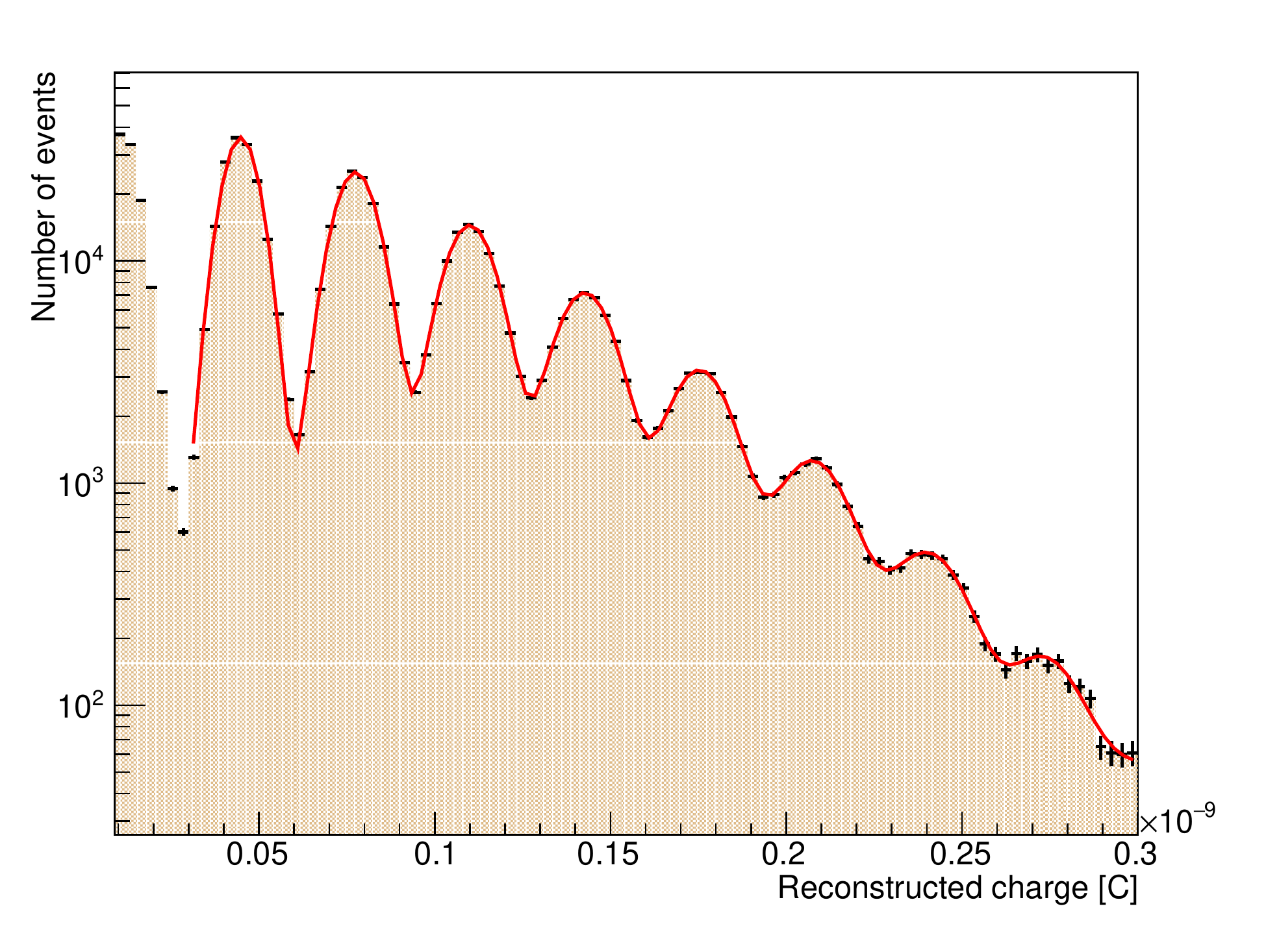}   
\end{center}
\caption{Reconstructed collected charge in the SAC. 
The distribution was approximated with the sum of eight gaussians, 
with parameters determined from a free fit.}
\label{fig:sac-charge}
\end{figure}
The reconstructed charge distribution 
is shown with points in Fig. \ref{fig:sac-charge}. The 
peaks correspond to 
1,2, ..., n
electrons impinging the 
detector. 
A fit to the collected charge with the sum of gaussians plus a constant 
was performed,
\begin{equation}
 N(x) = c + \sum_{i=1}^n N_i \times \exp\left( \frac{-(x - q_i)^2}{2\sigma_i^2} \right), 
\end{equation}
where the $q_i$ is the charge  corresponding to $i$ electrons 
in the detector, $\sigma_i$ is the charge resolution and $N_i$ are normalization factors.
All the parameters were left free, resulting in a
25 parameter fit, shown with red line in Fig.  \ref{fig:sac-charge}. 
The data and the model exhibit good agreement.

\begin{figure}[hptb]
\minipage{0.48\textwidth}
\begin{center}
\includegraphics[width=8cm]{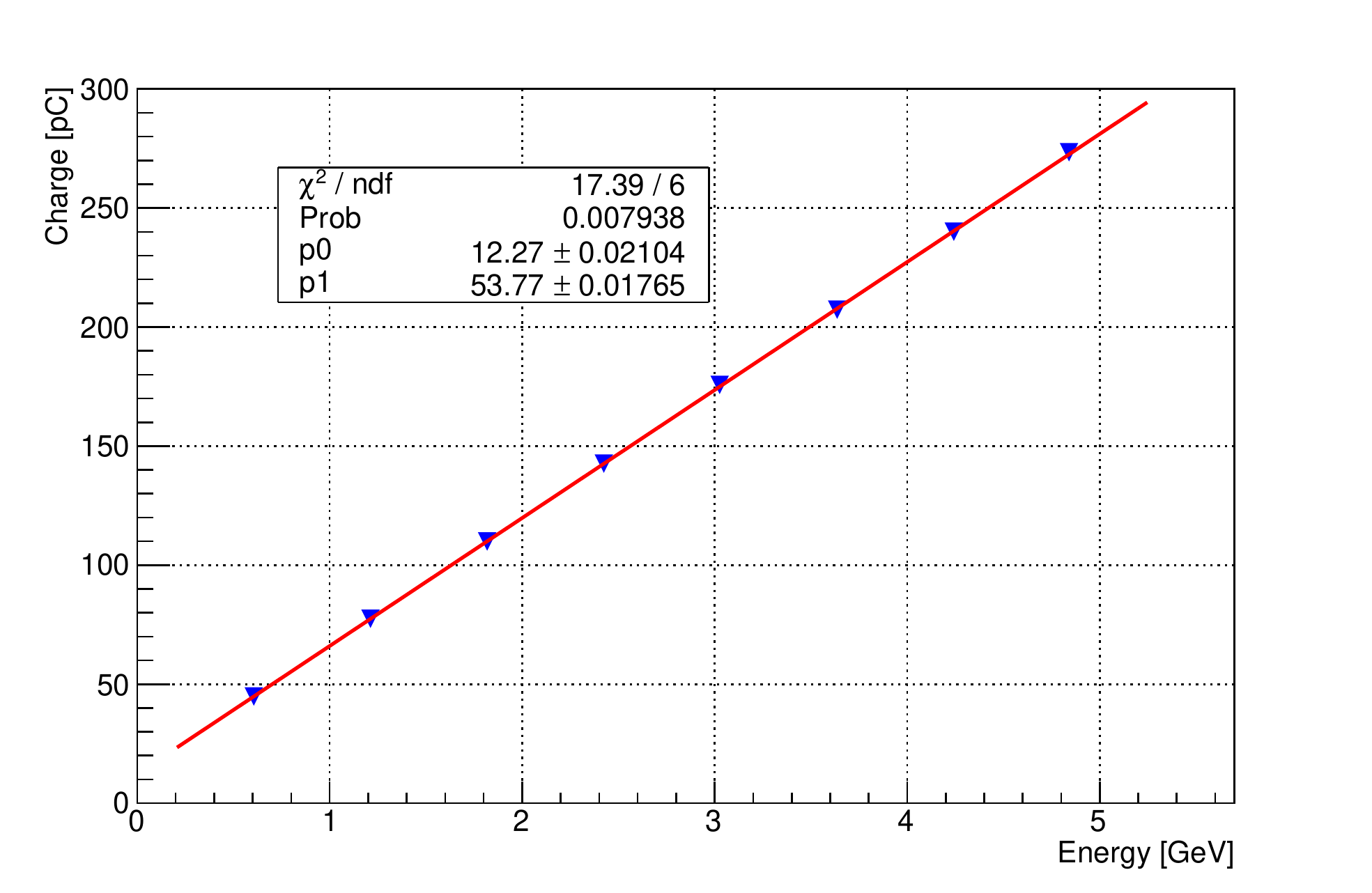}   
\end{center}
\caption{Relation between the reconstructed charge in the SAC and the deposited energy.
The linear fit of the points is also shown.}
\label{fig:sac-linearity}
\endminipage \hfill
\minipage{0.48\textwidth}
\begin{center}
\includegraphics[width=8cm]{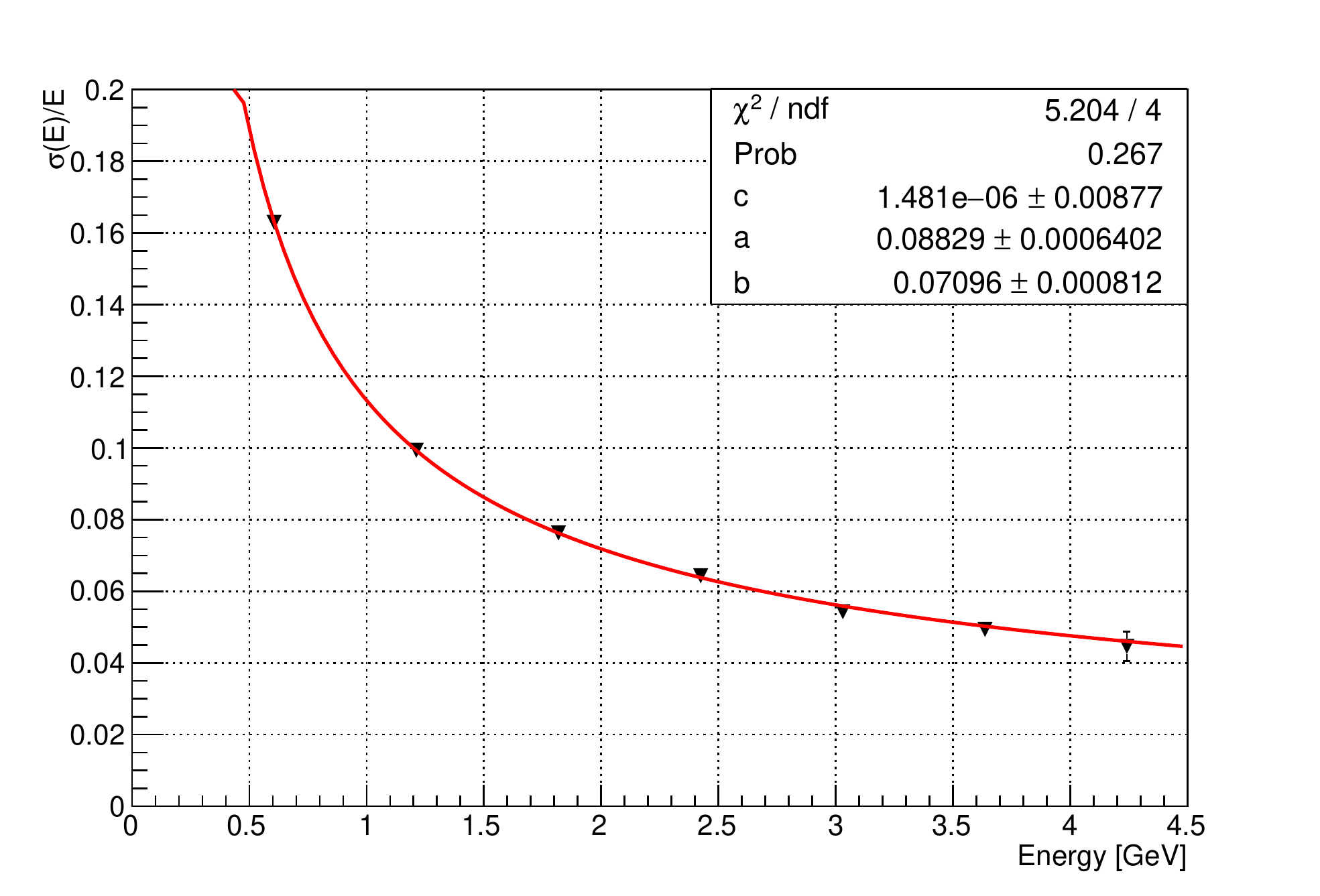}   
\end{center}
\caption{Energy resolution of the SAC as a function of the deposited energy (points)
approximated with the resolution function.}
\label{fig:sac-eres}
\endminipage \hfill
\end{figure}

The center of each of the gaussians 
allows to calibrate the relation between the collected charge 
and the deposited energy, as shown in Fig. \ref{fig:sac-linearity}. 
The response of the detector was verified to respect 
the linearity up to 4 GeV.

The dependence of the width of the gaussians 
normalized to the peak position (i.e. the energy resolution,  $\sigma(E)/E$ ) 
 as a function of the deposited energy is shown in Fig. \ref{fig:sac-eres}. 
It was parametrized, as 
\begin{equation}
 \sigma(E)/E = a/\sqrt{E} \oplus b/E \oplus c,
\end{equation}
where the energy is measured in GeV. The free parameters
were determined by fitting the points in Fig. \ref{fig:sac-eres}, 
to obtain $a$ = (8.8 $\pm$ 0.1)\%, $b$ = (7.1 $\pm$ 0.1)\%, and $c$ = (0 $\pm$ 1)\%. 


\section{Efficiency measurement}

The average number of the electrons per bunch 
was chosen to be close to 1. However, in the individual 
bunches the actual number of electrons varied from 0 to $\sim$10. 
To control the electron multiplicity, the signals from the two 
scintillator paddles were used. The collected charge for each of them 
was reconstructed on an event by event basis in the same way as 
was done for the SAC PMTs. 

\begin{figure}[hptb]
\minipage{0.48\textwidth}
\begin{center}
\includegraphics[width=8cm]{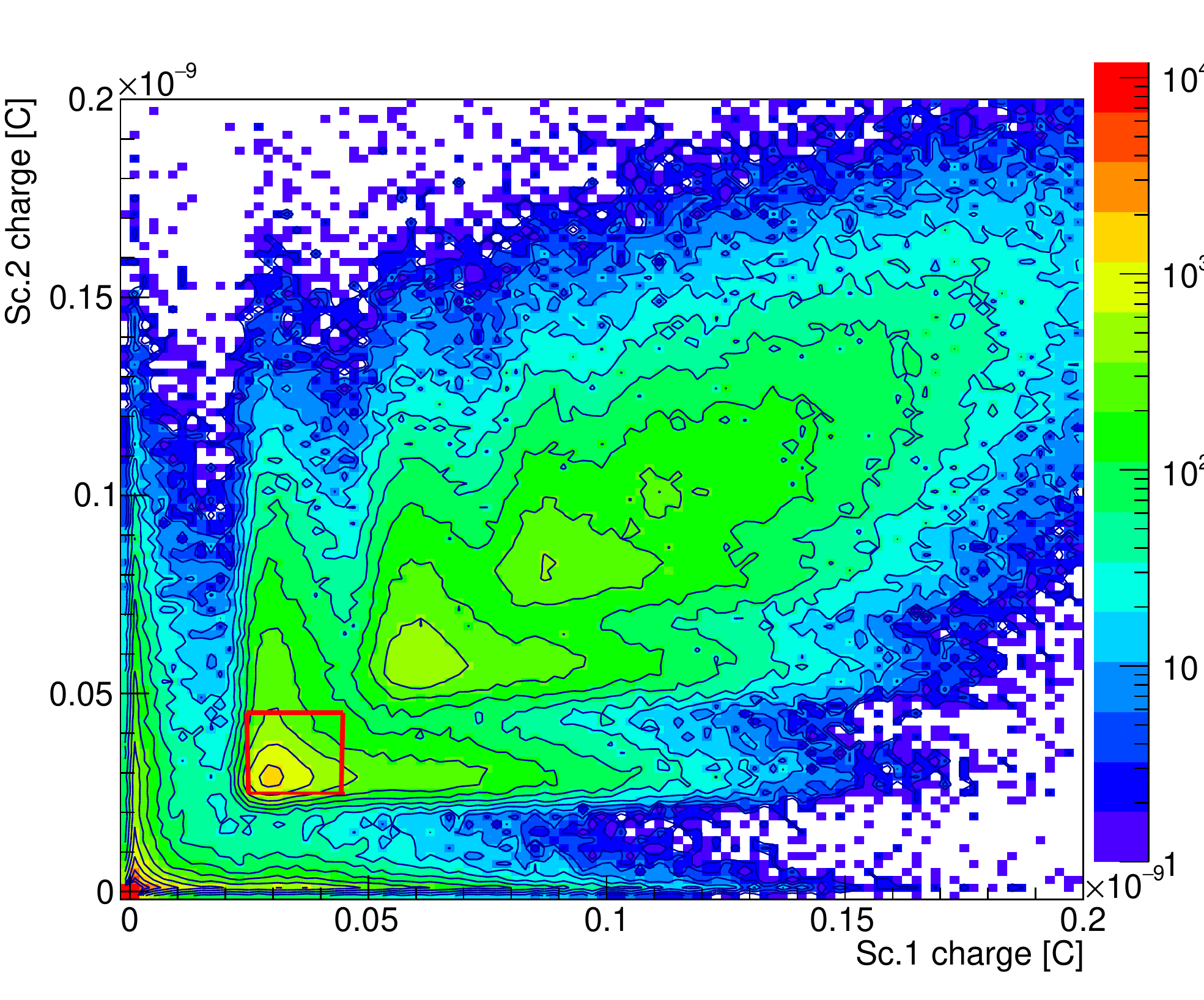}   
\end{center}
\caption{Reconstructed charge in the scintillator paddles, used to select single electron events.}
\label{fig:trigger-charge}
\endminipage \hfill
\minipage{0.48\textwidth}
\begin{center}
\includegraphics[width=8cm]{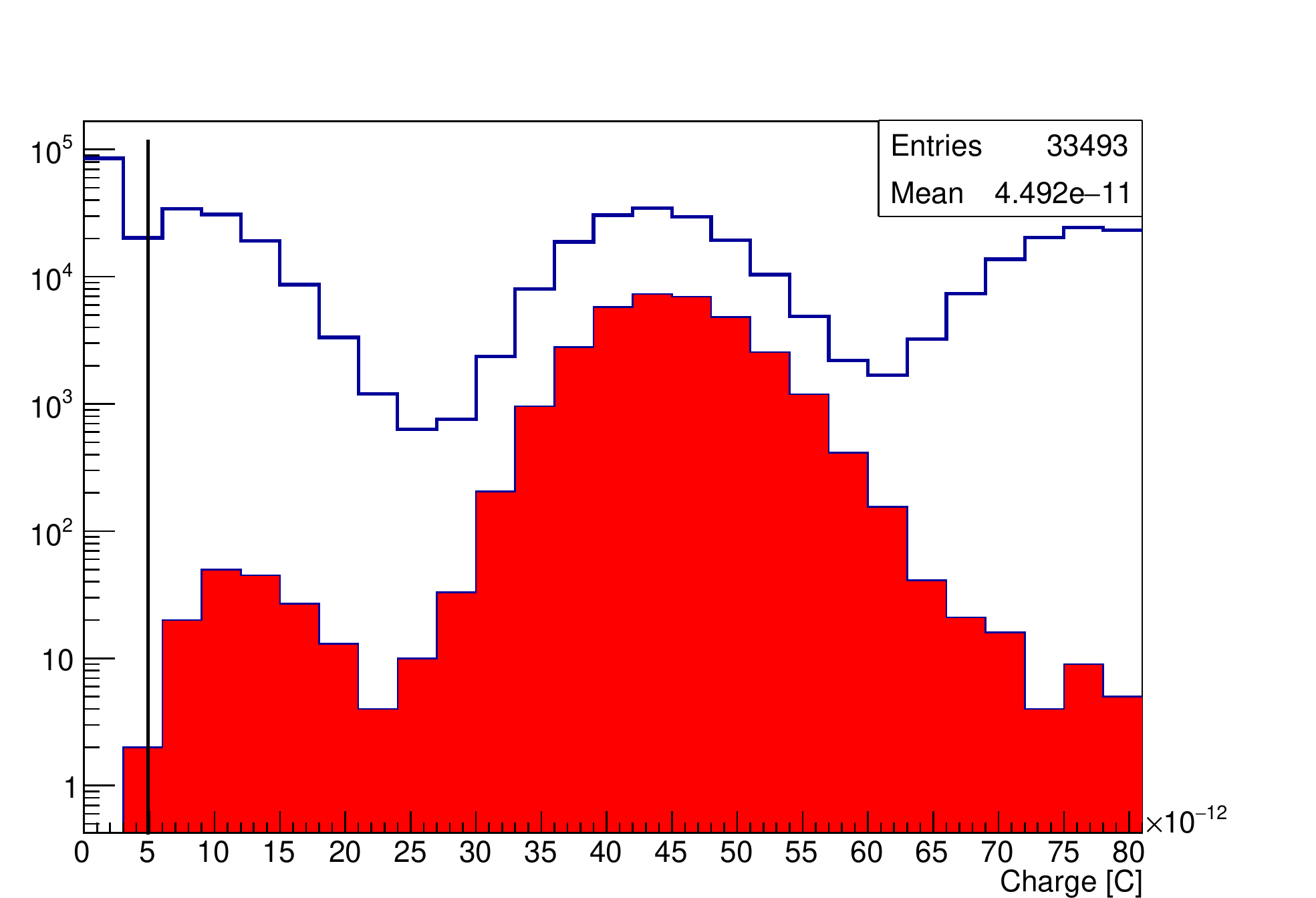}   
\end{center}
\caption{Reconstructed charge in the SAC without (white histogram) 
and with (filled histogram) the requirement of the single electron trigger. The 5 pC cut is shown with vertical line}
\label{fig:sac-eff}
\endminipage \hfill
\end{figure}

The reconstructed charge in Sc.1 versus the reconstructed charge in Sc.2 
is shown in Fig. \ref{fig:trigger-charge}. 
The islands of events with high population with equal charge in Sc.1 and Sc.2
correspond to the same number of electrons traversing both scintillators. 
A rectangular region, shown in red in Fig.  \ref{fig:trigger-charge} 
and defined as  25~pC~$\leq$~$q$(Sc.1),~$q$(Sc.2)~$\leq$~45~pC, 
was used to tag single-electron events. 

Out of the total sample, 33493 events were selected as 
single-electron candidates.
Their reconstructed charge in the SAC 
is shown as the filled histogram in Fig. \ref{fig:sac-eff}. 
A deposit of a 606 MeV energy in the SAC corresponded to $\sim$ 45 pC 
collected charge.
Most of the events cluster under the single electron peak;
however some of them exhibit lower deposited energy.

The number of events with reconstructed charge 
below 20 pC (on the left of the one electron peak)
was 161.   
By assuming that 
all they are undetected in SAC events, the inefficiency for detecting 
606~MeV electrons computes as $\eta = (5 \pm 0.4) \times 10^{-3}$. 
All tagged events exhibit reconstructed charge
that is above the pedestal, which peaks at zero 
as can be seen in Fig. \ref{fig:sac-eff}. 
The pedestal is defined as the reconstructed charge when no electron hits the SAC.
Setting a tentative cut of 5 pC as a minimal detectable charge, no event was found inefficient. 
This results in an estimation of the inefficiency of the 
SAC to detect 606 MeV electrons of  $\eta < 6\times10^{-5}$, at 90\% confidence level. 
The origin of the events between 5 pC and 20 pC was traced back to 
the impure tagging and the fact that there was no additional shielding employed to 
protect both the tagging paddles and the SAC itself. 
Low energy photons and
scraping/escaping low energy electrons could therefore reach the tagging paddles and the SAC and 
contaminate
the event sample used for 
the efficiency estimation. 
The measured value of $\eta = 5 \times 10^{-3}$  
is thus
interpreted as an 
upper limit on the SAC inefficiency to detect 606 MeV electrons. 

\section{Conclusion}

The upgraded Small Angle Calorimeter for the 
NA62 experiment was tested at the Frascati BTF. 
The detector showed good linearity in response to the deposited energy.
The energy resolution was measured to be 
$\sigma(E)/E = 8.8\%/\sqrt{E} + 7.1\%/E$,  
with $E$ in GeV, with a constant term consistent with zero.
In addition, the inefficiency for detecting 600 MeV particles was 
determined to be less than $5 \times 10^{-3}$, 
where the limit on the method comes from  
accidental activity in the BTF experimental hall. 
Currently, the SAC is part of the NA62 experimental complex and serves as an efficient photon veto.

\section*{Acknowledgements}

We warmly thank the BTF team, P. Valente, B. Buonomo, and L. Foggetta, for the 
excellent support and the useful discussions during the data taking.
We would like also to thank L. Litov for the permission to work with the SAC prototype.

\end{document}